\documentclass[aps,pre,twocolumn,amsfonts,amssymb,amsmath,showpacs,floatfix, superscriptaddress]{revtex4}

\usepackage{natbib, textcomp} 
\usepackage{amsfonts, amsmath,amssymb}
\usepackage{graphicx}
\graphicspath{{figs/}}

\newcommand{\hs}{\hspace{6pt}}
\newcommand{\eps}{\epsilon}
\newcommand{\dd}{\partial}

\newcommand{\uu}{\mathbf{u}}

\newcommand{\VV}{\mathbf{V}}
\newcommand{\WW}{\mathbf{W}}

\newcommand{\SSS}{\mathbf{S}}

\newcommand{\tuu}{\tilde{\mathbf{u}}}
\newcommand{\OO} {\mathcal {O}}

\newcommand{\la}{\lambda}

\newcommand{\HP}{\hat{\mathbf{P}}}

\newcommand{\HL}{\hat{\mathbf{L}}}

\newcommand{\bF}{\mathbf{F}}

\newcommand{\RR}{\mathcal{R}}

\newcommand{\nn}{\nonumber}

\newcommand{\bP}{\mathbf{P}}

\newcommand{\bY}{\mathbf{Y}}

\newcommand{\braket}[2]{\langle #1 \mid #2 \rangle  }

\newcommand{\bsub}{\begin{subequations}}
\newcommand{\esub}{\end{subequations}}

\newcommand{\etal}{\textit{et al.}}

\newcommand{\SR}{\mathbf{S}^{\mathrm{R}}}
\newcommand{\SdR}{\mathbf{S}^{\mathrm{dR}}}
\newcommand{\dxspiral}{\textsc{dxspiral}} 

\newcommand{\diag}{\mathrm{diag}}

\newcommand{\grad}{\mathrm{grad}}
\newcommand{\tens}[1]{\bar{\bar{#1}}}

\begin{document}

\title{Drift Laws for Spiral Waves on Curved Anisotropic Surfaces}
\author{Hans Dierckx}
\affiliation{Department of Mathematical Physics and Astronomy, Ghent
  University, 9000 Ghent, Belgium}
\author{Evelien Brisard}
\affiliation{Department of Mathematical Physics and Astronomy, Ghent
  University, 9000 Ghent, Belgium}
\author{Henri Verschelde}
\affiliation{Department of Mathematical Physics and Astronomy, Ghent
  University, 9000 Ghent, Belgium}
\author{Alexander V. Panfilov}
\affiliation{Department of Mathematical Physics and Astronomy, Ghent
  University, 9000 Ghent, Belgium}

\date{\today}

\begin{abstract}

Rotating spiral waves organize spatial patterns in chemical, physical and biological excitable systems. Factors affecting their dynamics such as spatiotemporal drift are of great interest for particular applications. Here, we propose a quantitative description for spiral wave dynamics on curved surfaces which shows that for a wide class of systems, including the BZ reaction and anisotropic cardiac tissue, the Ricci curvature scalar of the surface is the main determinant of spiral wave drift. The theory provides explicit equations for spiral wave drift direction, drift velocity and the period of rotation. Depending on the parameters, the drift can be directed to the regions of either maximal or minimal Ricci scalar curvature, which was verified by direct numerical simulations. \end{abstract}

\maketitle

\section{Introduction} Spiral waves of excitation have been observed in diverse chemical, biological and physical systems \cite{Winfree:1972, Zhabotinsky:1971, Jacubith:1990, Lechleiter:1991, Allessie:1973}.  
They organize spatial patterns of excitation and underly important processes such as morphogenesis of a social amoeba \cite{Siegert:1992, Nettesheim:1993}, some forms of neurological disease \cite{Gorelova:1983} and cardiac arrhythmias \cite{Allessie:1973, Gray:1995}. In many cases, the dynamics of spiral waves is of great interest because it determines the overall behavior of the system. One of the most important aspects of the dynamics is the spatiotemporal drift of spiral waves. The drift of spirals can determine the type of cardiac arrhythmia \cite{Gray:1995}; it has also been observed in the BZ reaction \cite{Agladze:1987, Maselko:1989}, CO oxidation on a Pt surface \cite{Jacubith:1990} and biological morphogenesis \cite{Nettesheim:1993}.  
Currently, several sources of spiral drift have been identified, including tissue heterogeneity \cite{Rudenko:1983, Panfilov:1991b, TenTusscher:2003, Dierckx:2009}, an external electrical field \cite{Steinbock:1992, Biktashev:1994b, Henry:2004}, spatially varying anisotropy \cite{Panfilov:1993c, Panfilov:1995, Berenfeld:1999, Wellner:2000, Davydov:2004} and surface curvature \cite{Davydov:2000}. The latter is highly relevant as most of real excitable media have complex geometries which may include curved domain boundaries, e.g. the walls of human atria are very thin and have a complex curved shape. In sufficiently thin slabs of excitable medium, we recently showed \cite{Dierckx:2012} that dynamics of spiral activity is essentially two-dimensional, and may therefore be modeled as a surface or monolayer of grid elements \cite{Blanc:2001, Jacquemet:2003}. Although the most detailed numerical models of human atria include heterogeneous electrophysiology and multiply layered fiber structure in specific wall regions, \cite{Vigmond:2001, Seemann:2006, Aslanidi:2011}, the magnitude and direction of spiral wave drift due to the wall shape and anisotropy alone has not yet been quantified. 

Essential questions regarding drift of spiral waves on a surface are: What determines the drift direction and velocity of the drift? How is spiral wave drift affected by anisotropy of the medium? Can the parameters of drift be predicted from general properties of 2D spiral waves? Some of these questions were addressed previously in the kinematic approach \cite{Zykov:1996, Davydov:2000, Davydov:2004}. However, the kinematic approach used there is valid only for spirals with a large core, i.e. where front-tail interactions are absent. 
However, curved surfaces with anisotropy were never studied before, despite their usefulness for cardiac applications.  

Here, we propose a theory of spiral wave drift on curved anisotropic surfaces based on a gradient expansion around the spiral wave solution. We derive equations for drift of a spiral wave on a surface of arbitrary shape with anisotropy.  
We show that the drift velocity is given by the gradient of the so-called Ricci curvature scalar (RCS) of the surface, which arises  as the generalization of the Gaussian curvature of a surface. 
The coefficients in our equation for drift velocity are explicitly obtained from the properties of the two-dimensional spiral wave solution in an isotropic planar medium using response functions \cite{Keener:1988, Biktasheva:1998, Henry:2002, Biktasheva:2009}. 
\begin{figure}[b]
\raisebox{3cm} {a)} 
\includegraphics[width = 0.21 \textwidth]{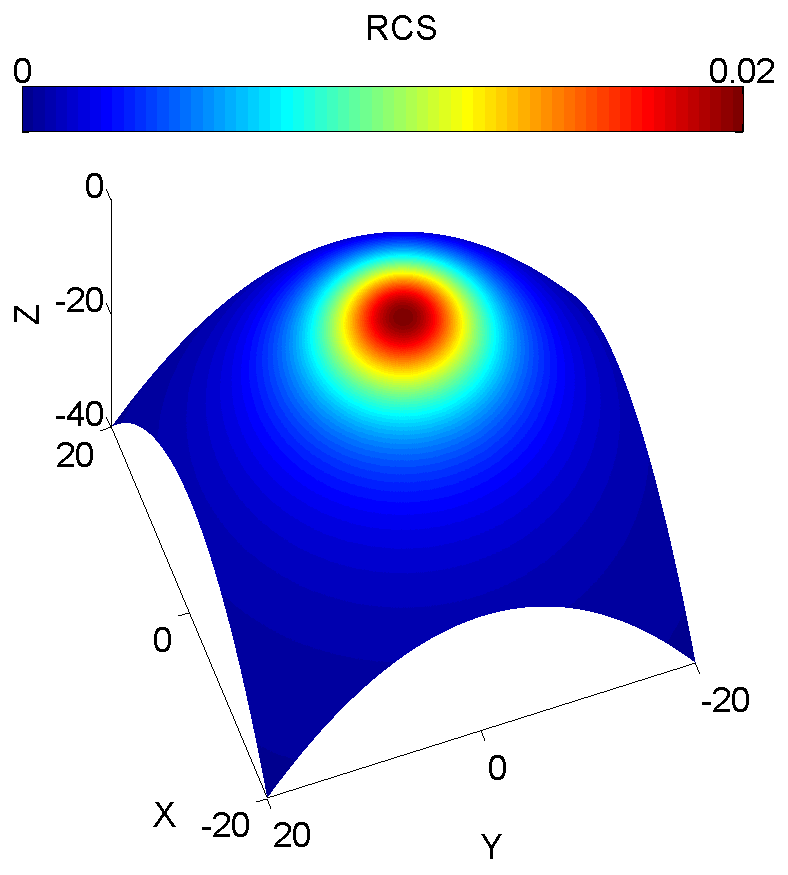}
\raisebox{3cm} {b)} 
\includegraphics[width = 0.21 \textwidth]{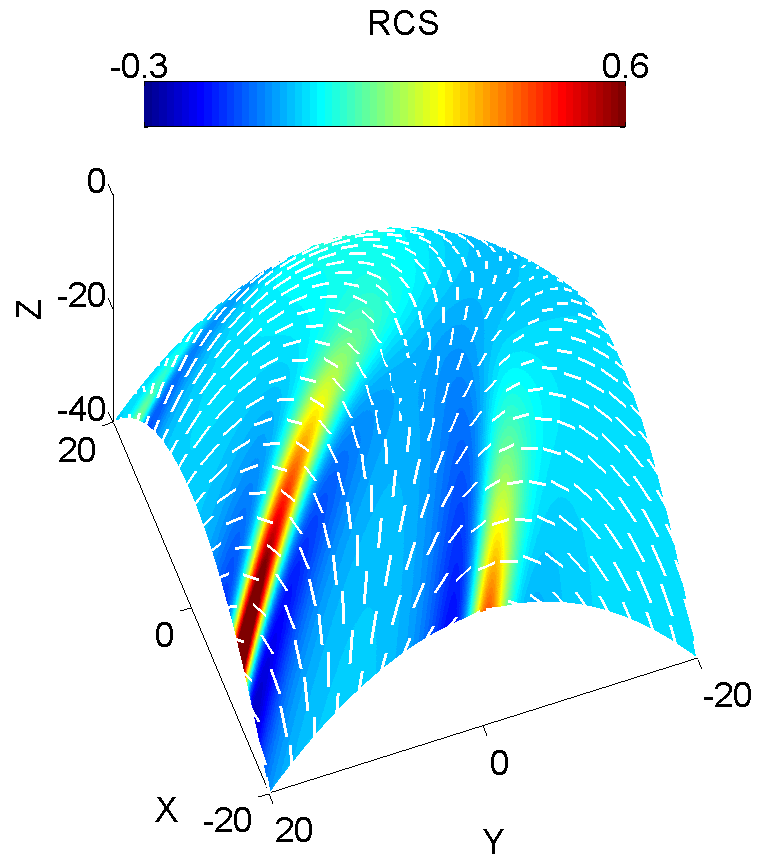}\\
\caption{Curved anisotropic surfaces colored according to their Ricci curvature scalar. (a) Paraboloid $z=-0.05(x^2+y^2)$ with isotropic diffusion. (b) Same paraboloid with anisotropic diffusion ($D_L/D_T=4$) with fiber angle $\alpha = (\pi/40)(x+y)$.  White bars indicate the local fiber direction.
\label{fig:geomsricci}}
\end{figure}
Interestingly, depending on parameters the drift can be directed to regions of the surface with either lowest or highest RCS. As the spatial distribution of the RCS can be easily computed (see paragraph \ref{sec:metric}), the proposed theory can predict the regions which will attract or repel spiral waves in each particular situation. Note that for anisotropic diffusion, the extrema of RCS do not necessarily coincide with the places of extremal surface curvature, an example of which is provided in Fig. \ref{fig:geomsricci}.  We verify our theory by a direct comparison with numerical simulations and show that the derived equations predict with high accuracy the trajectories of spiral wave drift on curved surfaces with significant anisotropy.

\section{Analytical methods and results}

\subsection{Reaction-diffusion equation on a curved surface with anisotropic diffusion} 
We start from the reaction-diffusion equation (RDE) in terms of Cartesian coordinates $x^i$. Anisotropy is built in through the diffusion tensor $D^{ij}$, whose eigenvalues are proportional to the squared conduction velocities along the local material axes:
\begin{equation}\label{RDEcart}
    \dd_t \uu(\vec{r},t) = \dd_i\left( D^{ij}(\vec{r}) \dd_j \mathbf{P} \uu(\vec{r},t) \right) + \mathbf{F}(\uu(\vec{r},t)).
\end{equation}

\noindent This PDE describes how a state vector $\uu$ of the system changes due to local processes $\bF(\uu)$ and anisotropic diffusion. The constant, dimensionless matrix $\bP$ allows to exclude some state variables from diffusion. We now derive the analogue of Eq. \eqref{RDEcart} on curved surfaces with isotropic or anisotropic diffusion.

\subsubsection{Isotropic diffusion on a surface.} Any  smooth surface can be parameterized as $x^i(s^A)$, $(i\in\{1,2,3\}, A\in\{1,2\})$, where the  $s^A$ form a curvilinear coordinate system. The gradient and divergence operators in the diffusion term of Eq. \eqref{RDEcart} should therefore be expressed using the metric tensor $G_{AB}$ that is induced on the surface by the coordinate transform $x^i \rightarrow s^A$, with components $G_{AB}= \dd_A x^i \delta_{ij} \dd_B x^j$. In case of isotropic diffusion $(D^{ij} = D_0 \delta^{ij})$, RD systems are thus described by \cite{Davydov:2000}
\begin{align}
\dd_t \uu &= \frac{D_0}{\sqrt{G}} \dd_A \left( \sqrt{G} G^{AB} \dd_B \bP \uu \right) + \bF(\uu). \label{RDE2}
\end{align}

\subsubsection{Anisotropic diffusion on a surface}
When diffusion is anisotropic, the diffusive current for a given diffusion tensor $\tens{D}$ equals $\vec{\mathbf{J}} = - \tens{D} \cdot \grad\, \bP \uu$. 
Transformation to surface coordinates brings $\mathbf{J}^A = - D^{AB} \dd_B \bP \uu$, where
\begin{equation}
D^{AB} = \dd_i s^A D^{ij} \dd_j s^B. \qquad (i,j\in\{1,2,3\}) \label{Dtrans}
\end{equation}
Taking the divergence will give $\frac{1}{\sqrt{G}} \dd_A \left( \sqrt{G} D^{AB} \dd_B \bP \uu \right)$ as the diffusion term for curved, anisotropic surfaces. From Eq. \eqref{Dtrans}, however, it follows that $ \det(D^{AB})= \det(G_{AB}))^{-1} \det(D^{ij})$. Here, $\det(D^{ij})$ is the product of the diffusivities along the local material axes and in most of the cases assumed to be constant \cite{Dierckx:2009}. 
Thus, following work by Wellner \etal\ \cite{Wellner:2002} as in \cite{Verschelde:2007, Young:2010} we can write for curved surfaces with anisotropic diffusion that 
\begin{align}
\dd_t \uu &= \frac{1}{\sqrt{g}} \dd_A \left( \sqrt{g} g^{AB} \dd_B \HP \uu \right) + \bF(\uu),  \label{RDE3}
\end{align}
with the metric tensor still given by the matrix inverse of the diffusion tensor, albeit in surface coordinates $s^A$:
\begin{align} \label{metricsurf}
  g_{AB} &= D_0 (D^{-1})_{AB} =  D_0 (D^{-1})_{ij} \dd_A x^i \dd_B x^j.
\end{align}
The constant factor $D_0$ has been included to make $g_{AB}$ dimensionless as is common in physics; we also define $\HP = D_0 \bP$.  

\subsection{Derivation of the drift equations}

The derivation of the laws of spiral wave drift occurs in three steps, similar to some of our previous works on wave fronts \cite{Dierckx:2011} and three-dimensional scroll waves \cite{Verschelde:2007, Dierckx:2009}. The first step is to choose the simplest curvilinear coordinates based on the problem's geometry. Next, we expand the RDE in those coordinates. Finally, we take inner products with response functions to obtain the equation of motion. 

\subsubsection{Riemann Normal Coordinates}
To describe spiral waves on a surface, we introduce Riemann normal coordinates \cite{MTW} around the spiral's rotation center at $t=0$. In these coordinates, the radial lines from the origin are geodesics of the space considered. 
Such coordinates explicitly reveal how curvature affects the metric in the region close to the origin, in terms of the Riemann curvature tensor $R_{ABCD}$. This tensor contains second order spatial derivatives of the metric \cite{MTW}; its trace is the Ricci curvature scalar $\RR$ mentioned above. The RCS can be computed from the metric tensor using the Christoffel symbols $\Gamma^A_{BC}$ \cite{MTW}:
\begin{align}
 \Gamma^A_{BC} &=  \frac{g^{AD}}{2}\left( \dd_B g_{CD} + \dd_C g_{BD} - \dd_D g_{BC} \right), \label{defR}\\
\RR &= g^{BC} \left(  \dd_A \Gamma^A_{BC} - \dd_C \Gamma^A_{AB} + \Gamma^A_{AD} \Gamma^D_{BC} - \Gamma ^A_{CD} \Gamma^D_{AB} \right). \nn 
\end{align}
We shall associate an order $\la$ to each spatial differentiation of the metric tensor, as we are working in the regime of slowly varying anisotropy and small Gaussian curvature $K_G$ of the surface in comparison with the spiral's core size. Hence, the fact that Riemann normal coordinates are locally Euclidean can be written as $g_{EF} = \delta_{EF} + \OO(\la^2)$. The explicit expansion for the metric tensor in the Riemann normal coordinates can be found in differential geometry textbooks (e.g. \cite{MTW}):
\bsub \label{gexp} \begin{align}
 &g_{EF}(\rho^1, \rho^2) = \delta_{EF} + \frac{1}{3} R_{EABF}(0,0) \rho^A \rho^B+  \label{metricRNC}\\
&\frac{1}{12} \left[ \dd_C R_{EABF}(0,0) + \dd _C R_{FABE}(0,0) \right] \rho^A \rho^B \rho^C + \OO(\la^4). \nn 
\end{align} 
In this expression, the Riemann curvature components and its derivatives were evaluated at the center of rotation of the spiral wave solution, where $\rho^1=\rho^2=0$. 

To find the metric components with upper indices, a matrix inversion is performed:
\begin{align}
 &g^{EF}(\rho^1, \rho^2) = \delta^{EF} - \frac{1}{3} R^{E\hs \hs F}_{\hs AB} (0,0) \rho^A \rho^B - \frac{1}{12} \times  \label{metricRNCcon}\\
& \left[ \dd_C R^{E\hs \hs F}_{\hs AB}(0,0) + \dd_C R^{F \hs \hs E}_{\hs AB}(0,0) \right] \rho^A \rho^B \rho^C + \OO(\la^4). \nn
\end{align} \esub
Since we are dealing with two spatial dimensions only, we can use the identity
\begin{equation}
R_{ABCD} = (\RR/2) (g_{AC} g_{BD} - g_{AD} g_{BC})
\end{equation}
whence, omitting terms of $\OO(\la^4)$, 
\bsub \label{Rprop} \begin{align}
  R_{EABF}(0,0) &= \frac{\RR(0,0)}{2} \left( \delta_{EB} \delta_{AF} - \delta_{EF} \delta_{AB} \right) \nn \\
& = \frac{\RR(0,0)}{2} \eps_{EA} \eps_{BF} \\
 \dd_C R_{EABF}(0,0) &= \frac{\dd_C \RR(0,0)}{2} \left( \delta_{EB} \delta_{AF} - \delta_{EF} \delta_{AB} \right) \nn \\ &= \frac{\dd_C \RR(0,0)}{2} \eps_{EA} \eps_{BF} 
\end{align} \esub
After substituting these relations in Eq. \eqref{metricRNCcon}, we may write the diffusion term of Eq. \eqref{RDE3} as
\begin{eqnarray}
\label{DTapp}
&& \frac{1}{\sqrt{g}} \dd_A \left( \sqrt{g} g^{AB} \dd_B \HP \uu \right) \\
 &=& \left( \frac{1}{\sqrt{g} } \dd_A \sqrt{g} \right) g^{AB} \dd_B \HP \uu + \dd_A \left( g^{AB} \dd_B \HP \uu \right) \nn \\
 &=& \frac{1}{2} g^{EF} \dd_A g_{EF} g^{AB} \dd_B \HP \uu + \dd_A \left( g^{AB} \dd_B \HP \uu \right) \nn \\  
&=& \dd^A \dd_A \HP \uu  - \RR(0,0) \SSS^{\rm R} - \dd^A \RR(0,0) \SSS^{dR}_A + \OO(\la^4),\nn 
\end{eqnarray}
 with the $\SSS^{\rm R}$ and $\SSS^{\rm dR}_A$ given by 
\begin{eqnarray}
 \SR  &=& \frac{1}{6} \left( \HP { \dd^2_\theta \uu_0} 
- \HP {r \dd_r \uu_0} \right), \label{SRdR}  \\
\SdR_A &=& - \frac{1}{6} \HP \rho_A {r \dd_r \uu_0} 
+ \frac{1}{12} \HP \rho_A {\dd_\theta^2 \uu_0}  
+ \frac{1}{24} \HP r^2 {\dd_A \uu_0}.   \nn
\end{eqnarray} 
Herein, $\dd_\theta = \eps_A^{\hs B} \rho^A \dd_B$, $r\dd_r = \rho^A \dd_A$ and $r^2 = \delta_{AB} \rho^A \rho^B$. The terms $\SSS^{\rm R}$ and $\SSS^{\rm dR}_A$ are the sources of spatiotemporal drift of spiral waves on the surface. 

\subsubsection{Expansion around the unperturbed spiral wave solution}

In the presence of source terms, the exact solution $\uu(\rho^A, t)$ can be approximated by an unperturbed spiral solution $\uu_0(\rho^A)$, i.e. 
\begin{equation} 
\uu = \uu_0 +\tuu, \label{utuu}
\end{equation}
where $\tuu= \OO(\la^2)$. 
Our approach thus relies on a linearization around the unperturbed spiral wave solution $\uu_0$. The associated linear operator
\begin{equation}
 \HL = \HP \Delta + \omega_0 \dd_\theta + \bF'(\uu_0)
\end{equation}
has critical eigenmodes for each Euclidean symmetry of the RDE \eqref{RDEcart}. Therefrom, one can prove the existence of one rotational and two rotational Goldstone modes $\dd_\theta \uu_0, \dd_x \uu_0, \dd_y \uu_0$ which are sometimes written $\VV_{(n)}, \{-1,0,1\}$ in a complex basis \cite{Biktashev:1994b, Biktasheva:2006, Biktasheva:2009}: 
\begin{align}
 \VV_{(0)} &= - \dd_\theta \uu_0, & 
\VV_{(1)} &= - \frac{1}{2} \left( \dd_x \uu_0 - i \dd_y \uu_0 \right).
\end{align} 
With respect to the inner product  
\begin{equation}
\braket{\mathbf f}{\mathbf g} = \int_{\mathbb{R}^2} dS \mathbf{f}^H \mathbf{g}, \label{inner}
\end{equation}
one may define the adjoint operator $ \HL^\dagger = \HP^H \Delta - \omega_0 \dd_\theta + \bF'^H(\uu_0) $, which has critical eigenmodes 
\begin{align}
 \VV_{(0)} &= - \dd_\theta \uu_0, & 
\VV_{(1)} &= - \frac{1}{2} \left( \dd_x \uu_0 - i \dd_y \uu_0 \right),\\
 \WW^{(0)} &= - \bY^\theta, \nn & 
\WW^{(1)} &= - \left(\bY^x - i \bY^y\right).
\end{align} 
that are known as response functions \cite{Biktasheva:1998}. 

The response functions can be used to fix the decomposition \eqref{utuu} by demanding that 
\begin{align}
 \braket{\bY^\theta}{\tuu} &= 0, &
 \braket{\bY^x}{\tuu} &= 0, & 
 \braket{\bY^y}{\tuu} &= 0
\end{align}
for all times $t$. This condition lets the origin of the Riemann normal coordinates move along with the spiral wave's rotation center and rotate at the yet unknown rotation frequency $\omega$, as in \cite{Verschelde:2007}. The differentiation with respect to time $\tau$ in this moving frame will generate convection terms 
\begin{equation}
  \dd_\tau \uu = \dd_t \uu - \omega \dd_\theta \uu- \dd_t X^A \dd_A \uu.
\end{equation}
Adding time-derivative, reaction and diffusion terms, we finally obtain 
\begin{multline}
 \HL \tuu  -\dot{\tuu} + \dot{X}^A \dd_A\uu_0 + (\omega-\omega_0) \dd_\theta \uu_0 \\ = \RR \SR - \dd^A \RR \SdR_A + \OO(\la^4) \label{master}
\end{multline}
with source terms given by \eqref{SRdR}.

\subsubsection{Rotational and translational dynamcis\\ using response functions}

The components of spiral drift can be found by projecting Eq. \eqref{master} onto the  response functions; this procedure relies on the Fredholm alternative theorem. The response functions were observed to be strongly localized around the spiral wave's tip \cite{Biktasheva:2003, Henry:2002}. This property ensures that dynamics of the full spiral wave can be essentially captured by expansion in Riemann normal coordinates close to its tip. Our method first delivers the instantaneous laws $\omega - \omega_0= \braket{\bY^\theta}{ \SSS}$,  $\dot{X}^A = \braket{\bY^A}{\SSS} $ where the brackets refer to the inner product \eqref{inner}.

If the drift and spiral core radius are small compared to the distance over which curvature and anisotropy change, one may average over one rotation to find the net spiral wave drift as
\bsub \label{eom1} \begin{eqnarray}
  \dd_t \phi &=& \omega_0 + q_0 \RR + \OO(\la^4), \label{eom1rot}\\
\dd_t \vec{X}  &=& - q_1 \grad\,\RR - q_2 \vec{n} \times \grad\, \RR. \label{eom1trans}
\end{eqnarray} \esub
where $\vec{n}$ is a unit normal vector to the surface, the $\grad$ operator taken with respect to the metric \eqref{metricsurf} and
\bsub \label{qcoeff} \begin{align}
 q_0  &= \braket{\bY^\theta}{\SR}, & &\\
q_1 &= \frac{1}{2} \braket{\bY^A}{\SdR_A}, & 
q_2 &= \frac{1}{2} \eps^A_{\hs B} \braket{\bY^B}{\SdR_A}.
\end{align} \esub
The Eqs. \eqref{RDE3}, \eqref{eom1}-\eqref{qcoeff} are the main analytical results of this paper. 

In index notation, our law of motion \eqref{eom1trans} becomes
\begin{equation}
 \dot{X}^A = - q_1 g^{AB} \dd_B \RR - q_2 g^{-1/2} \eps^{BA} \dd_B \RR + \OO(\la^5).  \label{eomtransindex}
\end{equation}
This expression is particularly useful for practical calculations or numerical implementation.

\section{Analysis of the drift equations} 
The law \eqref{eom1trans} for spatial drift is strikingly similar to the law of motion of electrons in a solid material \cite{Kittel:SSP}, where $\dd_t {\vec{X}} = - \mu\, \grad\,  \phi$ under an electric field $\vec{E} = - \grad\, \phi$. We will thus henceforth call $q_1$ the `spiral mobility'.
For positive mobility $q_1$, the spiral wave will descend the  gradient, ending up in a locus of minimal RCS, while for negative $q_1$ the spiral will drift to the region with maximal RCS  value.
Spiral waves, however, exhibit also a second component of drift with coefficient $q_2$ which makes them drift under an angle $\tan^{-1} q_2/q_1$ with the direction of the gradient of RCS. However, $q_2$ cannot influence whether the spiral drifts to higher or lower RCS. If the sense of spiral rotation is reversed, $q_1$ remains the same, while $q_2$ switches sign. Since the RCS-induced drift contains third order spatial derivatives of the diffusion tensor, it is different from the metric drift in \cite{Dierckx:2009} due to variations in $\det(\mathbf{D})$. While the proportionality constants for the metric drift equal the filament tension \cite{Biktashev:1994}, we have found no simpler expression for the spiral mobility $q_1, q_2$. 

Let us now consider how the laws \eqref{eom1} apply to an isotropic diffusion system, as for example the BZ reaction. Here, $\RR/2 = R^{12}_{\ \,12} = K_G$, so the RCS is simply twice the Gaussian curvature of the surface. Eqs. \eqref{eom1} then confirm results obtained by Zykov \etal\ \cite{Zykov:1996, Davydov:2000} which were obtained in the kinematic approach. Our theory demonstrates that those results hold not only in the large core regime but for any stationarily rotating spiral wave. Also, it was suggested in \cite{Davydov:2000} that in the equal diffusion case ($\bP = \mathbf{I}$) the drift component parallel to the gradient of $\RR$ disappears, i.e. $q_1=0$. However, in our theory this is not the case, and $q_1$ can have any value depending on properties of the spiral wave. Our prediction is numerically confirmed in section \ref{sec:driftiso}.

For cardiac tissue and other anisotropic reaction-diffusion systems, Eq. \eqref{eom1trans} is the first analytical expression that captures the dynamics of spiral waves on a curved anisotropic surface. Spiral drift is shown to be related to the gradient of the RCS which depends both on curvature of the surface and tissue anisotropy. When the local direction of maximal diffusivity is known in the medium, we may consider a smooth anisotropic surface with constant principal diffusivities $D_L = d_L D_0, D_T = d_T D_0$, where $D_L \geq D_T > 0$. The respective eigenvectors of the diffusion tensor will be denoted $\vec{e_L}$, $\vec{e_T}$ ; in the context of cardiac tissue, $\vec{e_L}$ is known as the local fiber direction. In a local Euclidean frame one may then write that $ g^{AB} = d_T \delta^{AB} + (d_L-d_T) e_L^A e_L^B$. Hence, with $\vec{e_N}$ a unit normal to the surface, the drift law \eqref{eom1trans} can be written, with $\nabla$ the gradient operator on an isotropic surface, 
\begin{eqnarray}
  \dot{\vec{X}} &=&  -q_1 \left[  d_L \vec{e_L}\, (\vec{e_L} \cdot \nabla\RR) 
+  d_T\, \vec{e_T}\, (\vec{e_T} \cdot \nabla\RR) \right]\nn \\
&& \qquad - q_2 \sqrt{d_L d_T}\, \vec{e_N} \times \nabla \RR + \OO(\la^5). \label{eom2trans}
\end{eqnarray} 
For strongly anisotropic tissue ($D_L \gg D_T$), the drift will thus appear to occur almost along the local fiber direction, as long as $|q_1| \approx |q_2|$. 

To further elucidate the link between $\RR$ and $\vec{e_L}$, we compare $g_{AB}$ with the metric $G_{AB}$, which would have been present if the diffusion on the surface had been isotropic. In local Euclidean coordinates $p^a$ around a given point of the surface, one has
\begin{align}
 g^{AB} &= \frac{\dd q^A}{\dd p^a} \frac{D^{ab}}{D_0} \frac{\dd q^B}{\dd p^b},&
 G^{AB} &= \frac{\dd q^A}{\dd p^a} \delta^{ab} \frac{\dd q^B}{\dd p^b}. 
\end{align}
When necessary, the subscripts $._{(g)}$ and $._{(G)}$ will written to denote which metric is used when calculating a quantity. Notably, the RCS that appears in the laws of spiral motion \eqref{eom1}, \eqref{eom2trans}, is in fact $\RR_{(g)}$. Aided by surface coordinates where $\dd_{q_1}\vec{r} = \vec{e_L}$, we have obtained the decomposition 
\begin{align}
  \RR_{(g)} &= d_T \RR_{\rm shape} + (d_L-d_T) \RR_{\rm aniso}.\label{Rdec}
\end{align}
The two terms separately capture the extrinsic curvature and fiber structure of the surface:
\begin{eqnarray}
  \RR_{\rm shape} &=& \RR_{(G)} = 2 K_G, \label{R2} \label{Rdec2}\\
 \RR_{\rm aniso} &=& - 2 \frac{1}{\sqrt{G}} \dd^2_L\sqrt{G} 
= -2 \mathrm{div}_{(G)} \left[ \vec{e_L} (\mathrm{div}_{(G)} \vec{e_L}) \right].  \nn 
\end{eqnarray} 
Here $K_G$ is the Gaussian curvature, which in absence of anisotropy ($d_L=d_T$) is the only driving force of the spiral wave drift. Anisotropy dependent drift is a result of  the divergence in the fiber direction field $\vec{e_L}$ of the surface. To predict spiral drift, one may therefore either compute $\RR_{(g)}$ from $g_{AB}$ and its Christoffel symbols, or use the pair $G_{AB}$, $\vec{e_L}$ and Eqs. \eqref{Rdec}-\eqref{Rdec2}.

For anisotropic diffusion in a plane, the Gaussian curvature term vanishes, and in terms of the local fiber angle $\alpha(x,y)$, one may then show that
\begin{equation}
\RR_{shape}/2 = - (\dd_L \alpha)^2 + \dd^2_{LT} \alpha + (\dd_T\alpha)^2
\label{rexplane}
\end{equation}
in which $\dd_L$ and $\dd_T$ are directional derivatives along and across the local fiber direction.

A remark needs to be made here with respect to a particular fiber organization  known as chiral anisotropy\cite{Davydov:2004}, in which the fibers in the surface start at the origin and enclose a fixed angle $\alpha$ with the radial direction. Although one computes that $\RR_{(g)}=0$ everywhere except the origin, spiral drift and a rotation frequency shift have been observed \cite{Davydov:2004}. In the curved-space viewpoint, such systems have the same geometry as a cone which is not a Riemannian manifold in its apex. Spiral drift in chiral anisotropy thus falls outside the scope of our present study.

\section{Numerical validation}

\subsection{Numerical methods}

\subsubsection{Evaluation of the coefficients $q_0,q_1,q_2$ \\ using response functions \label{sec:dxsp}}

The numerical values for the coefficients $q_i\ (i=0,1,2)$ displayed in Fig. \ref{fig:qcoeffs} and listed as $q_i^{th}$ in the main text were acquired using an extension of the publicly available \dxspiral\ software in the following way. First, \dxspiral\ was used to generate a standard spiral solution $\uu_0$ for Barkley's reaction kinetics  \cite{Barkley:1991}, where $\uu = [u,\, v]^T$, $\bF = [f =\varepsilon^{-1} u (1-u)(u-(v+b)/a),\, u-v]^T$ and $\HP = \diag(1, D_v)$. The standard spiral solution $\uu_0$ was computed on a disc of radius $R=12.0$, using a polar grid with $N_r = 240$ and $N_\theta=128$. Thereafter, response functions $\bY^\theta$, $\bY^x$, $\bY^y$ were computed by the $dxlin.c$ routine, whose details are given in \cite{Biktasheva:2009}. Next, the overlap integrals \eqref{qcoeff} were evaluated using the trapezoid rule, with terms $\SR$, $\SdR_A$ given by Eq. \eqref{SRdR}. 

In this way we evaluated  the coefficients $q_0,q_1,q_2$ for the parameter set $a=1.1,\, b=0.19,\, \eps = 0.025$, $D_v=0$. The coefficients shown in Fig. \ref{fig:qcoeffs} for model parameter $a>1.1$ were found in steps of $0.025$ up to $a=1.4$ by calculating new solutions $\uu_0$ using the solution for the previous $a$ as an initial guess. For each value of $a$, the response functions and the overlap integrals were evaluated as above in order to find $q_0, q_1, q_2$ and $\gamma_1, \gamma_2$.  

\subsubsection{Finding the metric and RCS for a surface\\ with given shape and projected fiber angle \label{sec:metric}}

We verified the laws of motion \eqref{eom1} by direct numerical simulations in Barkley's model. In our examples, we consider
surfaces $z=f(x,y)$ in the domain $(x,y) \in [-L/2, L/2] \times [-L/2,L/2]$. The surface is thought to contain fibers in the direction $\vec{e_L}$ tangent to the surface, whose fiber angle is defined by $\tan \alpha = (\vec{e_L} \cdot  \vec{e_y}) / (  \vec{e_L} \cdot  \vec{e_x} )$. For a prescribed angle $\alpha(x,y)$, one therefore finds that $\vec{e_L} = N [\cos \alpha \vec{e_x} +  \sin \alpha \vec{e_y} + ( \cos \alpha \ \dd_x f +  \sin \alpha\  \dd_y f) \vec{e_z}]$; the factor $N(x,y)$ is chosen to give $\vec{e_L}$ unit length. Now, we assume for a moment that the three-dimensional space is filled with copies of such surface in the direction of $\vec{e_z}$. If diffusion along the local fiber direction occurs with diffusion coefficient $D_L=D_0 d_L$, while transverse diffusion has $D_T = D_0 d_T$, the three-dimensional anisotropy is determined by the tensor
\begin{align}
 D^{ij} = D_T \delta^{ij} + (D_L - D_T) e_L^i e_L^j.
\end{align}
In the curved-space approach, a metric with contravariant components $g^{ij} = D^{ij}/D_0$ is found in the three-dimensional space, with inverse $g_{ij}$. 
In our simulations, we chose to let the surface parameterization $s^A$, $A=1,2$ be $s^1 =x, s^2 = y$. From the transformation law $g_{AB} = \dd_A x^i g_{ij} \dd_B x^j$, one then finds
\begin{align}
 g_{11} &= g_{xx} + 2 g_{xz} \dd_x f + g_{zz} (\dd_xf)^2, \nn\\
 g_{22} &= g_{yy} + 2 g_{yz} \dd_y f + g_{zz} (\dd_yf)^2, \label{appgAB} \\
 g_{12} &= g_{xy} + g_{xz} \dd_y f + g_{yz} \dd_x f + g_{zz} \dd_xf \dd_y f. \nn 
\end{align}

Thereafter, the metric components $g^{AB}$ are found as the matrix inverse of $(g_{AB})$. From the coefficients $g_{AB}$, $g^{AB}$, it is straightforward to compute the RCS using the Christoffel symbols from Eq.~\eqref{defR}. 

\subsubsection{Forward evolution of the RDE on a surface\\ with anisotropic diffusion }

To check the validity and limitations of the theory, forward evolution of spiral waves was studied on curved anisotropic surfaces. Hereto, the reaction-diffusion equation \eqref{RDE2}  was discretized using the finite difference technique. 
With the purpose of studying generic surfaces whose shape is prescribed by $z=f(x,y)$ in Cartesian coordinates, the curvilinear coordinates on the surface were taken to be $s^1 = x, s^2 = y$. That is, the function $u^j(x,y)$ would provide a top view on the field of the j-th variable of the spiral wave. A rectangular grid with $dx=dy$ was taken. For the examples considered, we had $\dd_x f(0,0) = \dd_y f(0,0)=0$, such that the finest spatial grid on the surface was obtained in the origin. This value also determined the largest time step allowed in our explicit Euler scheme; we chose $dt = 0.9 dx^2 / (4 \max(D_L, D_T))$. 

The diffusion term in Eq. \eqref{RDE2} was discretized using a nine-point scheme, with the metric $g^{AB}$ the inverse of $g_{AB}$ from Eq. \eqref{appgAB}. For a given time $t$ and state variable label $j$, we took 
\begin{multline}
  \frac{1}{\sqrt{g}} \dd_A \left( \sqrt{g} g^{AB} \dd_B u^j(x , y ,t) \right) 
\approx \frac{1}{dx^2 \sqrt{g(x,y)} } . \\\sum \limits_{m,n \in \{ -1, 0, 1\} } C_{m,n}(x ,y) u^j(x+m dx ,y + n dx, t). 
\end{multline}
Simple finite differencing yields the coefficients $C_{m,n}(x ,y)$, which were only computed at the start of the simulation and then stored. With $h^{AB} = \sqrt{g} g^{AB}$, they are 
\bsub \begin{align}
C_{0,0}(x,y) =& - \sum_{m=\pm 1} h^{11}\left(x+m \frac{dx}{2}, y \right)  \\& -  \sum_{n=\pm 1} h^{22}\left(x, y+n \frac{dx}{2} \right), \nn \\ 
C_{m,0}(x,y) =& h^{11} \left( x+ m\frac{dx}{2}, y \right) \\
+& \sum_{n=\pm 1} \frac{mn}{4} h^{12} \left(x, y+n\frac{dy}{2} \right) , \quad (\mathrm{for}\ m=\pm 1) \nn \\
C_{0,n}(x,y) =& h^{22} \left( x, y+ n \frac{dx}{2} \right) \\
+& \sum_{m=\pm 1} \frac{mn}{4} h^{12} \left(x+m\frac{dy}{2},y \right),  \quad (\mathrm{for}\ n=\pm 1) \nn \\
C_{m,n}(x,y) =&\frac{mn}{4} h^{12} \left( x, y+ n \frac{dx}{2} \right) \\
+& \frac{mn}{4} h^{12} \left( x+ m \frac{dx}{2}, y \right) \qquad  (\mathrm{for}\ m,n=\pm 1). \nn 
\end{align} \esub

An overview of simulation parameters and grid size and resolution is presented in Tab.~\ref{tab:sims}. Before each simulation, a spiral wave was first created in a planar domain of larger size, same resolution and constant anisotropy equal to $g_{AB}$ at $(x,y)=(0,0)$. The midpoint of the circular tip trajectory was determined, such that the standard spiral wave solution could be copied and centered on a suitable position in the anisotropic curved surface. This method allowed to reduce the duration of the transient regime and the associated drift, and therefore brought more control of the initial spiral wave position.

\begin{table}[h]
 \begin{tabular}{rccc cccc c}
\hline
   Fig. & a &b& A & B & L & dx   & $D_v$ & $D_L$\\ 
\hline
 3 & 0.7 & 0.19&  0.1 & 0 & 40 & 0.1   &1 & 1\\
\hline
  4a & 1.3 & 0.19&  0 & $\pi$/40 & 30 & 0.1 & 0 & 4 \\
  4b & 1.1 & 0.19&  0 & $\pi$/40 & 30 & 0.1 & 0 & 4 \\
\hline
 5a (red) & 1.3 & 0.19&  0.5 & $\pi$/40 & 40 & 0.1 &  0 & 4 \\
 5a (yellow) & 1.3 & 0.19& 0.5 & 0 & 40 & 0.1 &  0 & 1 \\
 5b (red) & 1.1 & 0.19&  0.025 & $\pi$/80 & 80 & 0.1  &  0 & 4  \\
 5b (yellow) & 1.1 & 0.19& 0.025 & 0 & 80 & 0.1  &  0& 1  \\
\hline
\end{tabular}
\caption{Overview of simulation parameters. All simulations had $\eps =0.025$,$D_0=1$, $D_u=1$ and $D_T=1$ \label{tab:sims}}
\end{table}

\subsection{Numerical results}

\subsubsection{Predicted spiral mobility from response functions}

\begin{figure}[t]
\includegraphics[width = 0.45 \textwidth]{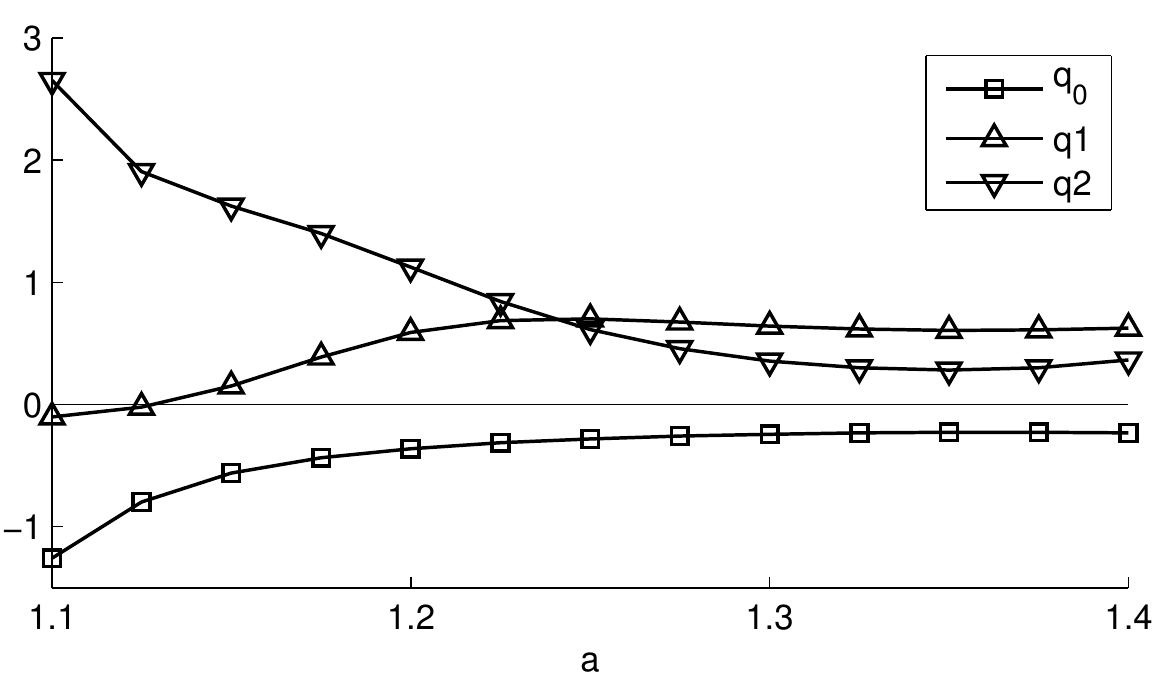} 
\caption{ Drift coefficients $q_0, q_1, q_2$ induced by Ricci scalar curvature in Barkley's model \cite{Barkley:1991} ($b=0.19, \eps = 0.025, \HP = \mathrm{diag}(1,0)$) for varying $a$. Signs of  $q_0, q_2$ for counterclockwise rotation. 
\label{fig:qcoeffs} }
\end{figure}

Fig. \ref{fig:qcoeffs} shows the dependency of the coefficients $
q_0, q_1, q_2$ as a function of the parameter $a$ which determines the excitability of the medium (the higher values of $a$ correspond to higher excitability). We see that the coefficient $q_1$ is positive for most values of $a$, indicating drift to the lower values of RCS. However, in a medium with low excitability, the spiral mobility $q_1$ can be also negative, making the spiral waves drift into the regions of higher RCS. For counterclockwise rotating spirals, the coefficient $q_2$ is always positive and slightly decreases with $a$, while the coefficient $q_0$ is negative and increases. In accordance with Eq. \eqref{eom1rot}, this explicitly shows that spiral waves rotate faster on sphere-like surfaces, as can be expected from the angular deficit, thereby extending the results of \cite{Zykov:1996} to non-uniformly curved surfaces with anisotropic diffusion. 

\subsubsection{Spiral wave drift on a paraboloid with isotropic diffusion \label{sec:driftiso} }

\begin{figure}[t]
\includegraphics[width = 0.4 \textwidth]{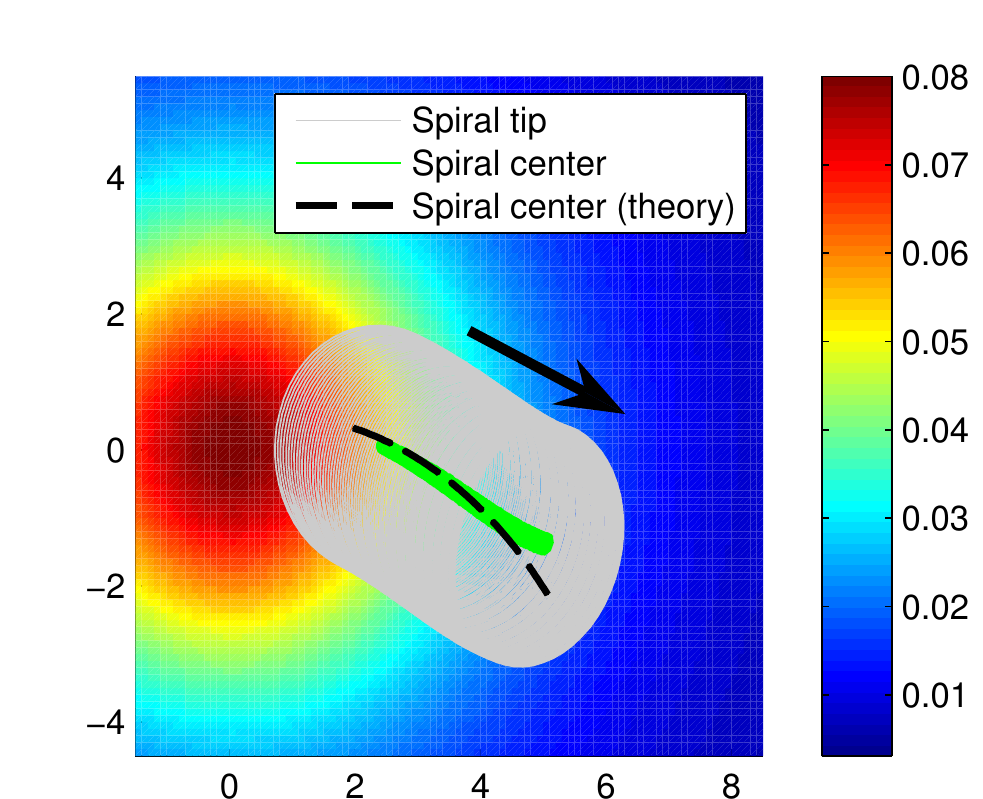}
\caption{Drift trajectory of a spiral wave on a paraboloid of revolution $z=0.1(x^2+y^2)$ in the equal diffusion case, showing a nonzero drift component along the gradient of the RCS. Colors indicate the RCS.
\label{fig:drifteq}
}
\end{figure}

As a first example, we studied the drift of a spiral wave on the paraboloid surface $z = -A(x^2+y^2)$ with equal diffusion, which has
\begin{equation}
  \RR= 2 K_G = 8 A^2 \left(1 + 4 A^2 (x^2+y^2)  \right)^{-2}.
\end{equation}
Barkley's model was used for the reaction kinetics, with $a=0.7$, $b=0.19, \eps = 0.025$ and $\HP= \diag(1,1)$. Although the kinematic approach in \cite{Davydov:2000} states that $q_1$ should vanish, it is clearly seen in 
Fig. \ref{fig:drifteq} that the spiral wave drifts away from the top, in accordance with Eqs.\eqref{eom1trans}-\eqref{qcoeff}, which yield $(q_1,  q_2) = ( 0.855, -0.386)$ for a counterclockwise spiral at the given model parameters.

This simple geometry allows to find analytically the spiral wave trajectory, which is drawn in black in Fig. \ref{fig:drifteq}. First, we introduce polar coordinates $(r, \theta)$ to exploit axial symmetry: $\dd_\phi \RR = 0$. Thus follows from Eq. \eqref{eom1}:
\begin{eqnarray}
 \dot{r} &=& - q_1 g^{rr} \dd_r \RR, \nn \\
 \dot{\phi} &=& - q_2 \frac{1}{\sqrt{g}} \dd_r \RR.
\end{eqnarray}
For the paraboloid $z=\pm Ar^2$, one finds $g^{rr} = (1+4A^2r^2)^{-1}$ and $g = r^2(1+4A^2r^2)$, whence
\begin{equation}
 \frac{d\phi}{dr}  =  \frac{q_2}{q_1} \frac{\sqrt{1+4A^2r^2}}{r}.
\end{equation}
Integration then brings
\begin{equation}
 \phi(r)  = \frac{q_2}{q_1} \left( \sqrt{1+4A^2 r^2} - \coth^{-1}  \sqrt{1+4A^2 r^2} \right) + C_1. \label{trajparab}
\end{equation}

\subsubsection{Drift of spiral waves in a plane with  anisotropic diffusion}

\begin{figure}[b]
\raisebox{5.5cm} {a)} \includegraphics[width = 0.37 \textwidth]{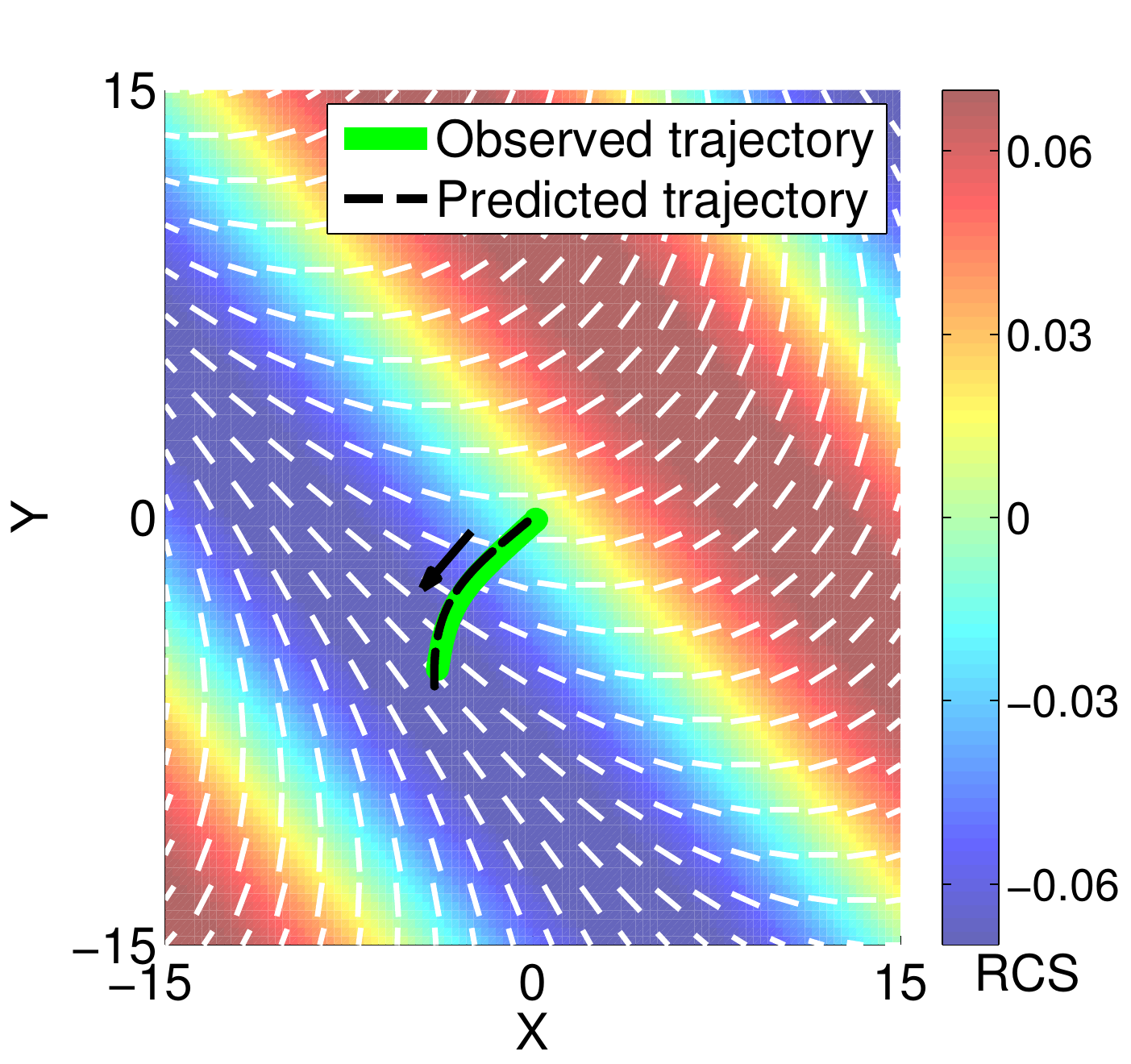} \\ 
\raisebox{5.5cm} {b)} \includegraphics[width = 0.37 \textwidth]{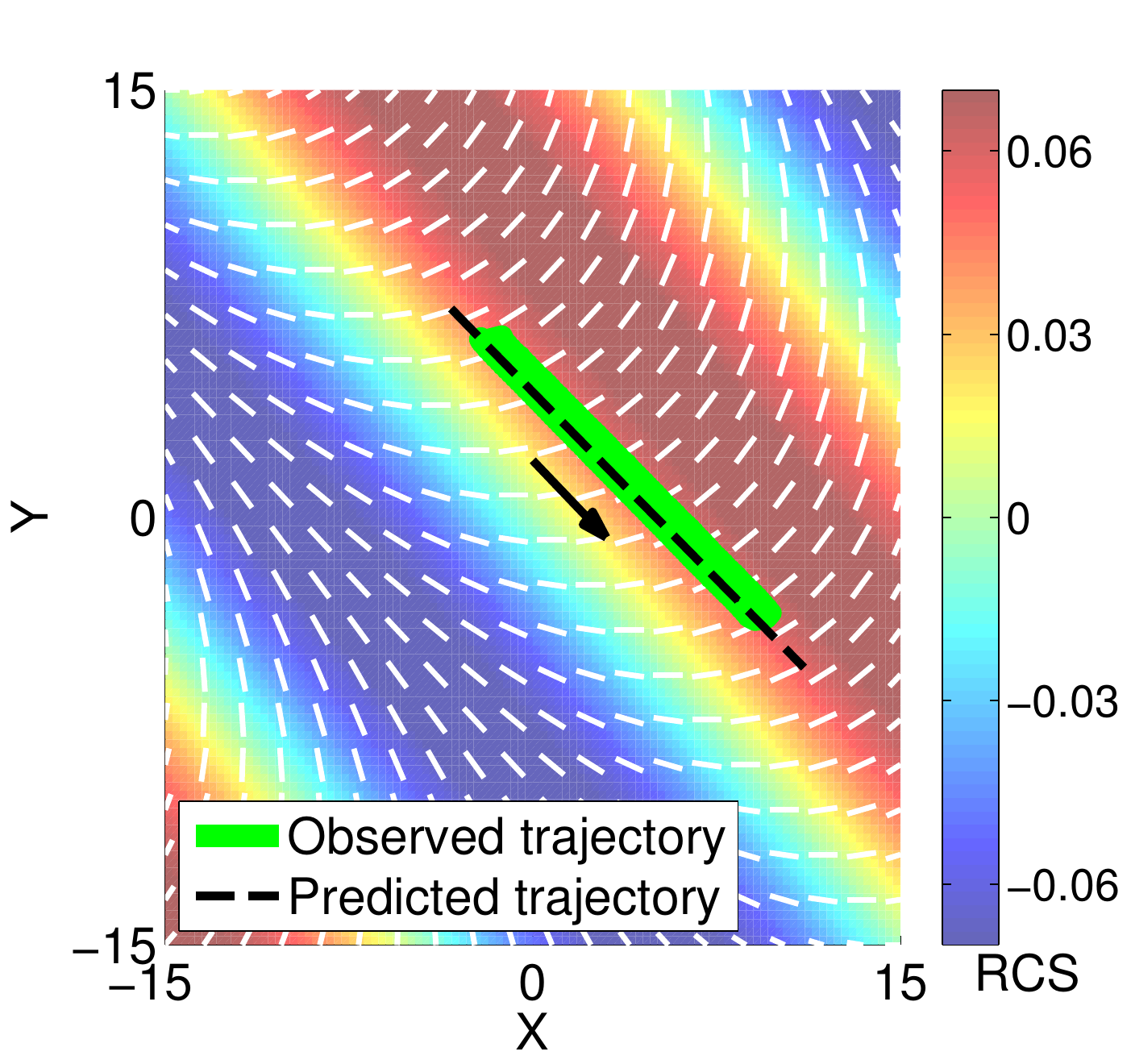}   
\caption{Attractive and repulsive sites for spiral waves in the anisotropic plane of Fig. \ref{fig:geomsricci}b, i.e. fiber angle $\alpha = (\pi/40)(x+y)$ and $D_L/D_T=4$. (a) Drift trajectories for Barkley's model as in Fig. \ref{fig:qcoeffs} with parameter $a=1.3$. (b)~Same for $a=1.1$. \label{fig:driftplanar}
}
\end{figure}

In our second numerical experiment, we considered an anisotropic plane with linear fiber rotation as in \cite{Rogers:1994}, i.e. $\vec{e_L} = \cos \alpha \vec{e_x} + \sin \alpha \vec{e_y}$ with fiber angle $\alpha(x,y) = B\,(x+y)$. For such anisotropy, a direct analytical calculation using Eq. \eqref{rexplane} gives 
\begin{equation}
 \RR = 4 (d_L-d_T) B^2 \sin 2 \alpha, 
\end{equation}
which is color-coded in Fig. \ref{fig:driftplanar}. 
The configuration does not possess isolated maxima or minima of the RCS: the local extrema are located along lines at $-\pi/4 $ to the $x$ axis.  The minima occur at the fiber angle $\alpha = -\pi/4 $, while the maxima are found where $\alpha = \pi/4 $. 

For this case too, an analytical spiral trajectory can be found. Going to coordinates $z=x+y$, $w=x-y$, the RCS is found to be independent of $w$. In these coordinates, Eq. \eqref{eom1} tells that
\begin{eqnarray}
 \dot{z} &=& - q_1 g^{zz} \dd_z \RR, \nn \\
 \dot{w} &=& - q_1 g^{wz} \dd_z \RR - q_2 \frac{1}{\sqrt{g}} \dd_z \RR,
\end{eqnarray}
whence
\begin{multline}
\frac{dw}{dz} = \frac{(d_L-d_T)\cos 2 \alpha(z)}{d_L + d_T + (d_L-d_T) \sin 2 \alpha(z)}\\ + \frac{q_2}{q_1} \frac{2 \sqrt{d_L d_T}}{d_L+d_T + (d_L-d_T) \sin 2 \alpha(z)} 
\end{multline}
This expression can be integrated to 
\begin{eqnarray}
&W(z) =  \ln \left[ (d_L-d_T)\sin 2 \alpha + d_L + d_T \right] &\\ 
  &\qquad + \frac{q_2}{q_1 B} \tan^{-1} \left[ \frac{(d_L+d_T) \tan \alpha + (d_L-d_T) }{2 \sqrt{d_L d_T}}   \right] + C_2. \nn &
\end{eqnarray}
The trajectory of the spiral wave's center in Cartesian coordinates is thereafter easily found as
\begin{align}
 x &= \frac{z+W(z)}{2},&  y &= \frac{z-W(z)}{2}. \label{trajplanar}
\end{align}
These relations are used for a prediction of the spiral trajectory in Fig. \ref{fig:driftplanar}.

To study both positive and negative mobility in this numerical example, we took $a=1.1$ or $a=1.3$, for which  Eqs. \eqref{qcoeff} respectively predict $(q_1, q_2) = (-0.102, 2.652)$ and $(q_1, q_2) = (0.643, 0.357)$ if the spirals rotate counterclockwise. We observe in Fig. \ref{fig:driftplanar}a that for positive mobility $q_1$ the spiral wave drifts towards minimal value of RCS, as predicted by our theory. In addition, we see a good correspondence of the real computed  trajectory (green) and the one predicted by Eq. \eqref{trajplanar} (black). For the negative mobility $q_1$ in Fig. \ref{fig:driftplanar}b, we observe only a small drift component towards the maximal value of RCS, as for this parameter value $|q_1| << |q_2|$. Here too, the theoretical (black) and computed (green) trajectories almost coincide. 

\subsubsection{Drift of spiral waves on a paraboloid surface \\ with anisotropic diffusion}

\begin{figure}[t]
\raisebox{6cm} {a)} \includegraphics[width = 0.38 \textwidth]{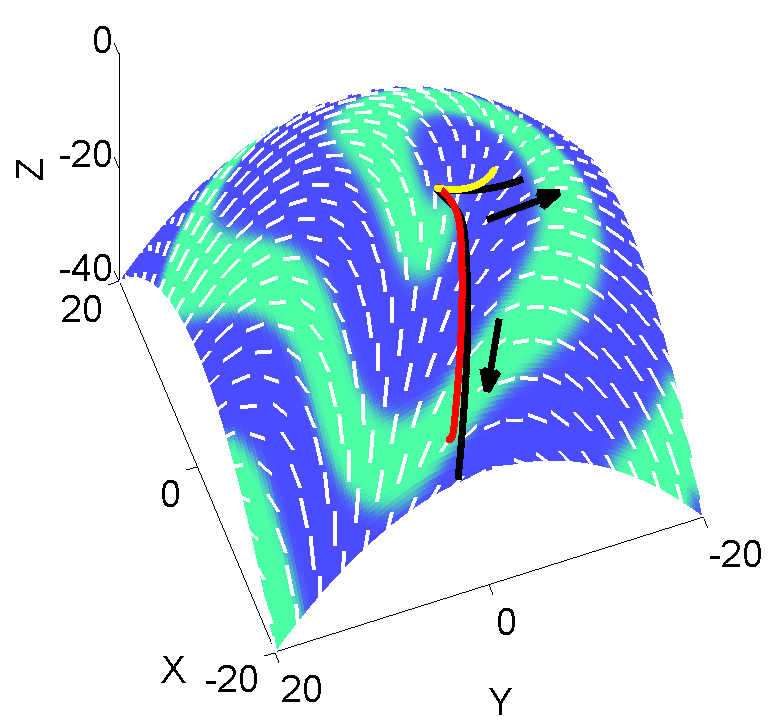}   \\ 
\raisebox{6cm} {b)} \includegraphics[width = 0.38 \textwidth]{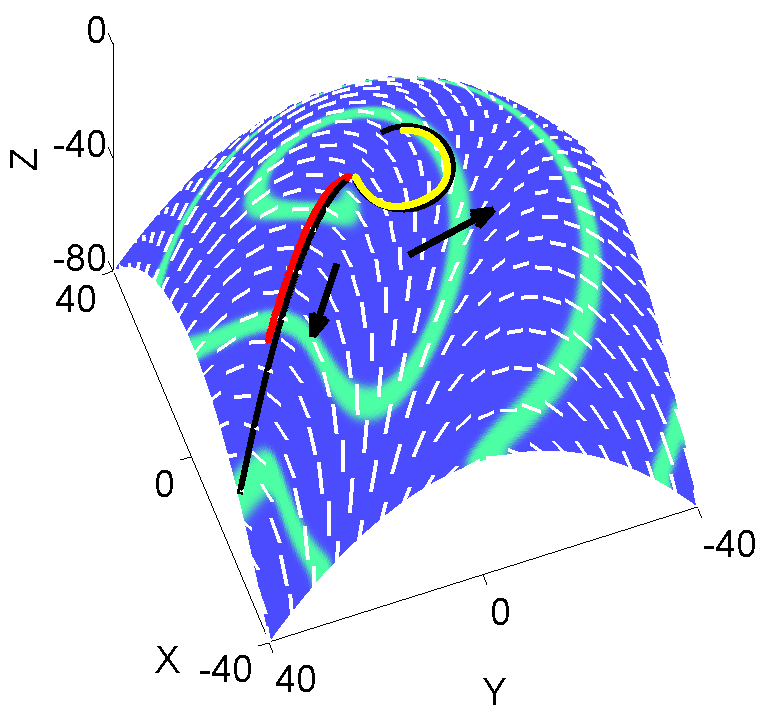}  
\caption{Spiral wave drift on a paraboloid surface; white bars indicate fiber direction. Drift trajectories for isotropic diffusion (yellow) and anisotropic diffusion as in Fig. \ref{fig:geomsricci}b (red) together with theoretical predictions (black). Barkley's model as in Fig.~\ref{fig:qcoeffs} and parameters $a=1.3, A=0.05, B=\pi/40$ (panel a) or $a=1.1, A=0.025, B=\pi/80$ (panel b). \label{fig:driftparab}
}
\end{figure}

In a third numerical experiment, the planar surface was replaced by the paraboloid $z=-A(x^2+y^2)$ with the same  anisotropic properties as in Fig.~\ref{fig:driftplanar}. Its RCS was calculated by numerically and already shown in Fig. \ref{fig:geomsricci}b. We see that the observed drift trajectory (Fig. \ref{fig:driftparab}, red lines) is in close agreement with the theoretical predictions obtained from numerical integration of the equation of motion \eqref{eomtransindex} (black). A good agreement between theory and experiment was reached for both cases of positive and negative $q_1$. We also provide trajectories for spiral wave drift in absence of anisotropy, i.e. only due to Gaussian curvature of the surface (yellow lines).  We see that for isotropic case the spiral indeed drifts away from the top of the paraboloid ($q_1>0$, Fig. \ref{fig:driftparab}a), or slowly towards it ($q_1<0$ , Fig. \ref{fig:driftparab}b), in accordance with the analytical trajectories \eqref{trajparab}.

\section{Discussion}

In this paper, we derived the laws of motion \eqref{eom1} for spiral waves on curved surfaces with anisotropic diffusion. Using the Fredholm alternative theorem, we showed that on such surfaces, the driving force for spiral wave drift is a gradient of the Ricci curvature scalar. This quantity is determined solely by the geometry of the surface; Eq. \eqref{rexplane} shows that it consists of two terms related to the either the shape or anisotropy of the surface. Thus, the current theory can be applied to a broad class of reaction-diffusion systems, exhibiting isotropic diffusion on a curved surface, anisotropic diffusion, or both. 

Although the correspondence to forward numerical simulations is excellent, several further steps can be undertaken to further increase the potential with respect to cardiac modeling. For example, spiral waves in more advanced models of cardiac tissue often exhibit quasi-periodic tip trajectories; this meandering motion will need to be included in the response function framework. 

Our present findings deal with two-dimensional vortices and thus do not consider effects of domain thickness at all. As for effects of anisotropy on two-dimensional spirals, we think that these effects may be substantial. One may recall one of the most cited papers on spiral waves in the heart \cite{Davidenko:1992}, which shows that even in simple two-dimensional preparations there is a substantial drift of spiral waves. As the preparations they used in their study were strongly anisotropic but otherwise homogeneous, the driving force of this drift is most likely the anisotropy of cardiac tissue. In real cardiac tissue, the setting will be three-dimensional and the effects of wall thickness and intramural fiber rotation may need to be added on top of our present study. 
Finally, it will be interesting to measure anisotropy-induced drift of spiral waves in a detailed ionic model of cardiac tissue, to estimate its magnitude in cardiac tissue.

\section{Conclusions} 
We have developed an asymptotic theory that predicts the drift of spiral waves on general curved surfaces with anisotropic diffusion. This drift is caused by a gradient of the Ricci curvature scalar, which encompasses both the shape and anisotropy of the surface. We determined the spiral mobility coefficients relating the gradient of  the Ricci curvature scalar and drift velocity using response functions. The analytical results were quantitatively confirmed by numerical simulations.

H.D. thanks the FWO Flanders for personal funding and computational infrastructure. The authors are grateful to Vadim Biktashev and Irina Biktasheva for helpful suggestions.


\end{document}